\documentclass[review,authoryear]{elsarticle}
\def\nat{\ref@Nature}
\def\apjs{\ref@ApJS}
\def\apjl{\ref@ApJ}
\def\apj{\ref@ApJ}
\def\aj{\ref@AJ}
\def\mnras{\ref@MNRAS}

\usepackage{graphicx}
\usepackage{epsfig}
\usepackage{geometry}                
\usepackage[parfill]{parskip}    
\usepackage{amssymb}
\usepackage{epstopdf}
\usepackage{rotating}
\usepackage{subfigure}
\title{Unusual Double-peaked Emission in the SDSS Quasar J093201.60+031858.7}
\author[spac]{R.S.~Barrows\corref{cor1}} 
\ead{rbarrows@uark.edu}
\author[spac,physuaf]{C.H.S.~Lacy} 
\ead{clacy@uark.edu}
\author[spac,physuaf]{D.~Kennefick} 
\ead{danilek@uark.edu}
\author[spac,physuaf]{J.~Kennefick} 
\ead{jkennef@uark.edu}
\author[physualr]{M.S.~Seigar} 
\ead{mxsseigar@ualr.edu}
\cortext[cor1]{Principal corresponding author}
\address[spac]{Arkansas Center for Space and Planetary Sciences, 202 Old Museum Building, University of Arkansas, Fayetteville, AR 72701}
\address[physuaf]{Physics Department, University of Arkansas, 825 West Dickson Street, Fayetteville, AR 72701}
\address[physualr]{Department of Physics and Astronomy, University of Arkansas at Little Rock, 2801 S. University Avenue, Little Rock, AR 72204}
\begin{document}
\maketitle

\section*{Abstract}
We examine spectral properties of the SDSS quasar J093201.60+031858.7, in particular the presence of strong blue peaks in the Balmer emission lines offset from the narrow lines by approximately 4200 km s$^{-1}$.  Asymmetry in the broad central component of the H$\beta$ line indicates the presence of a double-peaked emitter.  However, the strength and sharpness of the blue H$\beta$ and blue H$\gamma$ peaks make this quasar spectrum unique amongst double-peaked emitters identified from SDSS spectra.  We fit a disk model to the H$\beta$ line and compare this object with other unusual double-peaked quasar spectra, particularly candidate binary supermassive black holes (SMBHs).  Under the binary SMBH scenario, we test the applicability of a model in which a second SMBH may produce the strong blue peak in the Balmer lines of a double-peaked emitter.  If there were only one SMBH, a circular, Keplerian disk model fit would be insufficient, indicating some sort of asymmetry is required to produce the strength of the blue peak.  In either case, understanding the nature of the complex line emission in this object will aid in further discrimination between a single SMBH with a complex accretion disk and the actual case of a binary SMBH.  

\emph{Keywords}: accretion disks; galaxies: active; quasars: emission lines; black holes

PACS codes: 98.54.Aj, 98.54.Cm, 98.62Js, 98.62.Mw

\section{Introduction}
Quasars and active galactic nuclei (AGN) with spectra showing double-peaked emission lines are an interesting class of objects, though in many cases the origin of the double peaks is not fully understood.  Some of these spectral features have been examined in the context of emission from an accretion disk.  These objects are commonly known as double-peaked emitters (or disk emitters) and have been the subjects of a number of studies \citep{EH94,EH03,Strateva03}.  Double-peaked emitters are important objects to study because the double-peaked profiles provide information about the structure of the accretion disk around supermassive black holes (SMBHs).  The disk profiles are typically centered around Balmer emission line positions presumably because of photoionization and electron scattering in a  geometrically thin disk \citep{CH89}.  In the disk profiles, there is typically a peak on the red side and a peak on the blue side of the central emission line, though in some cases the peaks are very broad and shallow.  Furthermore, the blue peaks are typically stronger than the red peaks because of relativistic beaming.   

For many double-peaked emitters, a simple (circular, Keplerian) disk model provides a satisfactory model.  However, this sub-class of AGN contains a range of Balmer emission line profiles, and in some cases the fits require additional complexities (i.e. asymmetries) such as eccentricity in the disk or a hot spot on the disk surface.  In a few cases, only one peak appears to be offset from the narrow lines, and such objects have been interpreted as candidate binary or recoiling SMBHs because the spectra resemble multiple emission line regions.  These interpretations are important to consider because binary and/or recoiling SMBHs are expected to result from galaxy mergers \citep{MM04,Komossa06} and therefore many such cases should exist.  Additionally, they are believed to play a role in the evolution of the SMBH population \citep{Hopkins05}.  Therefore, there has been much interest in the possibility of identifying close binary SMBHs by locating double-peaked emission lines in the spectra of AGN and quasars \citep{Gaskell96,Zhou04}.  In these cases, determining the true physical origin of the double-peaks is difficult since they may be partly or entirely due to complex disk emission instead.  Interestingly, a few candidate binaries have been identified.  For example, the object SDSS J153636.22+044127.0 (from here on SDSSJ1536+0441) has been interpreted as a binary SMBH \citep{BL09}, a disk emitter \citep{Chornock10}, and as both \citep{TG09}.  Another candidate, SDSS J105041.35+345631.3 (from here on SDSSJ1050+3456) shows similar features and was first identified by \citet{Shields09} as a possible recoiling or binary SMBH.  A third candidate, SDSS J092712.65+294344.0 (from here on SDSSJ0927+2943), was first identified and interpreted as a recoiling SMBH \citep{Komossa08}, and more recently also as a binary SMBH \citep{Bogdanovic09b,Dotti09}.  The diversity of these interpretations indicates that accretion processes around a singe SMBH or in a binary SMBH are expected to produce a range of complex spectral signatures.  Results from simulations indicate that profiles of close binaries strongly resemble those of double-peaked emitters \citep{Bogdanovic08,Bogdanovic09a}; therefore, it is important to examine double-peaked profiles, particularly those with strong and distinct multiple peaks, as evidence for complex accretion processes that may be the result of a binary SMBH.  

Inspired by the above mentioned previous searches, and the results of those searches, we performed a similar search through the Sloan Digital Sky Survey (SDSS) archives to identify spectra with double-peaked emission lines at $z<0.89$; this redshift limit was chosen so that all spectra would include the H$\beta$ line.  All matches were individually inspected. In this paper, we describe the spectrum of a quasar, SDSS J093201.60+031858 (from here on SDSSJ0932+0318), that has a blue set of Balmer emission lines offset from the narrow lines by approximately 4200 km s$^{-1}$.  While the H$\beta$ line has a broad, red peak that is indicative of a double-peaked emitter, it has unusually strong blue H$\beta$ and H$\gamma$ peaks that are unusual among SDSS identified double-peaked emitters.  We examine similarities and differences between the features of SDSSJ0932+0318 and those of other unusual double-peaked emitters, particularly candidate binary SMBHs, to help determine the physical nature of the line emission.  Throughout the paper we adopt the cosmological parameters $H_o=71$ km s$^{-1}$ Mpc$^{-1}$, $\Omega_M=0.27$ and $\Omega_{vac}=0.73$.  

\section{Spectral Analysis}
\subsection{The Continuum}
The quasar spectrum we discuss was drawn from the fourth edition of the SDSS quasar catalogue \citep{Schneider07}.  It has SDSS point source magnitudes of $u=18.731$, $g=18.460$, $r=18.386$, $i=18.271$ and $z=17.945$.  Using the suggested conversion from \citet{Jester05},  $B=g-0.17(u-g)+0.11$, yields $B=18.39$, which is also corrected for a Galactic color excess of $E(B-V)=0.053$, taken from the NASA Extragalactic database following the model of \citet{Schlegel98}.  Converting to a bolometric luminosity \citep{Marconi04} and correcting for a stellar contribution of 33 percent \citep[the mean value from][]{EH03} yields $L_{BOL}=4.25\times10^{45}$ erg s$^{-1}$.  The spectrum was corrected for Galactic extinction using the IRAF task `deredden.'  Since disk emission is evident in the spectrum (figure 1), the line of sight to the central SMBH is probably not significantly obscured so we did not correct for internal extinction.  The continuum slope was modeled as a power law (f$_{\lambda}=\lambda^{-\alpha}$), and the best fit continuum slope has an index of $\alpha=2.05$. 

\begin{figure}
\centerline{\epsfig{figure=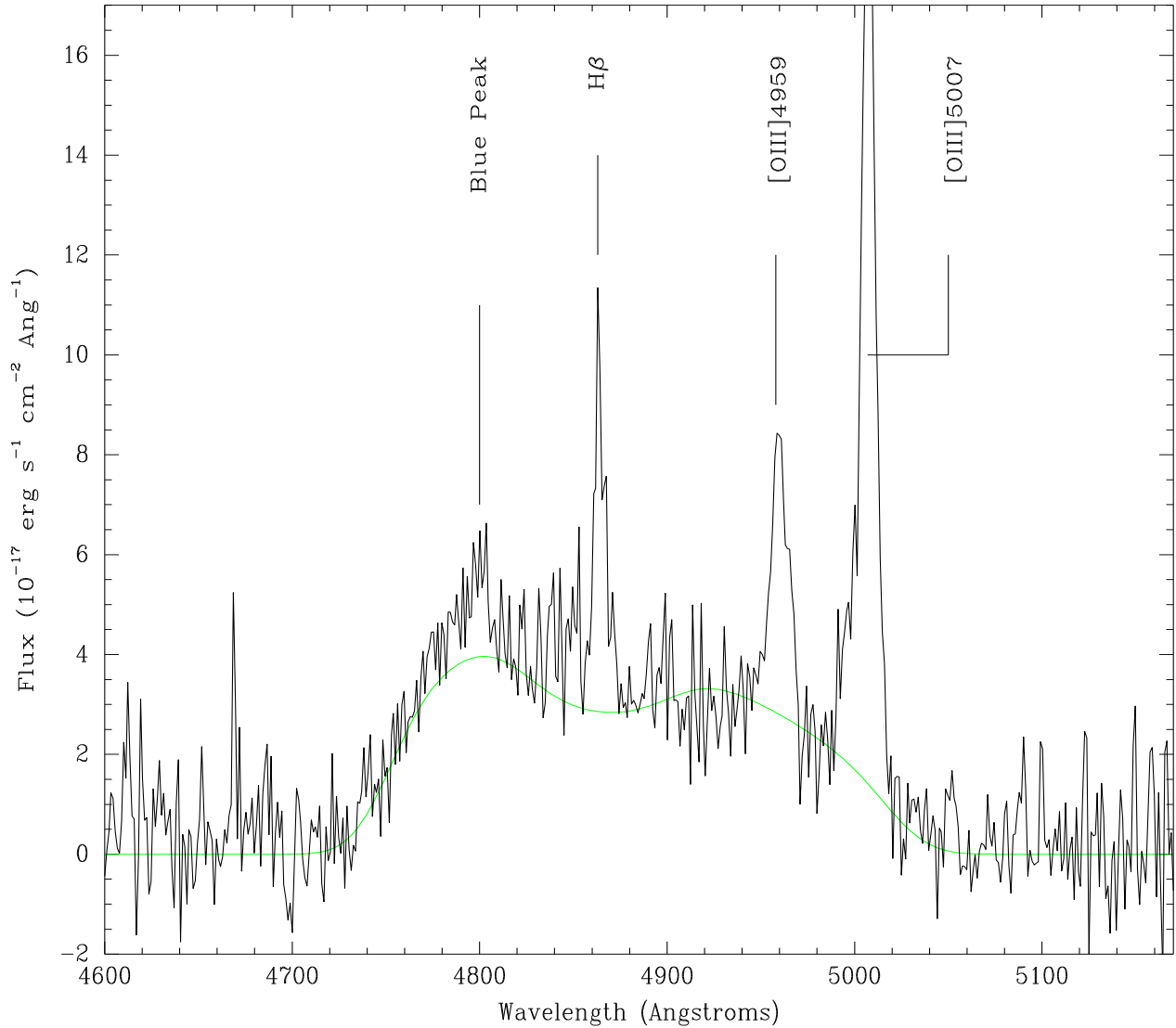,height=18cm,width=16cm,angle=270}}
\caption{H$\beta$/[OIII] complex at the rest wavelength with the disk profile overlaid.  The central wavelength of the disk model is set to the position of the narrow H$\beta$ line.  The parameters of the fit are as follows: $q=3$, $\sigma=1000$ km s$^{-1}$, $i=30^\circ$, $\xi_1=225$ $r_g$ and $\xi_2=2700$ $r_g$.  Notice the strong blue peak that is not adequately fit by the circular, Keplerian disk model.}
\label{Hb_wdisk}
\end{figure}

\subsection{The Disk Profile}
The H$\alpha$ line has been redshifted almost entirely out of the spectrum, therefore we primarily focus on the H$\beta$/[OIII] complex.  To fit the asymmetric line profile and to test the applicability of the disk-emitter scenario, we fit a disk profile to this region.  The disk model was centered at the redshift of the [OIII]5007 line based on the assumption that it represents the redshift of the central SMBH (discussed in section 3.2).  The disk model we used was adopted from \citet{CH89} and assumes a circular, Keplerian disk.  The assumption of circular shape is used since an accretion disk, in the absence of external influences, should be circularized due to the energy loss within the disk.  In the model, which has been widely used to model double-peaked profiles, the parameters are as follows: the local broadening of the line ($\sigma$), the inclination of the disk (\emph{i}), the inner and outer radii of the disk ($\xi_1$ and $\xi_2$) in gravitational units ($r_g$), and emissivity as a function of radius on the disk ($\epsilon=\xi^{-q}$) where \emph{q} is a dimensionless power law index.  In previous uses of this model, the power law index \emph{q} is usually constrained near 3.  Therefore, we have set $q=3$ to minimize the number of free parameters.  The local broadening tends to range from $\sigma\approx500 - 2000$ km s$^{-1}$, and we used a value of 1000 km s$^{-1}$ which is within that range and typical of disk model fits in the literature.  The inclination, inner radius and outer radius are therefore the three parameters that we adjusted to find the best fit, and each best fit was obtained by eye with the restriction that the fit be bounded by the observed emission line as was done in \citet{CH89}.  We fixed the inclination at three values, 15$^{\circ}$, 30$^{\circ}$, and 45$^{\circ}$, and varied the inner and outer radii to obtain the best fit for each inclination.  For 15$^{\circ}$, the best fit produces a sharp red peak that does not match the broad, shallow red peak that is observed.  In contrast, for 45$^{\circ}$, the best fit does not produce the extended red wing that is blended with the [OIII] emission lines.  We adopted an inclination of \emph{i}=30$^{\circ}$ and inner and outer radii of $\xi_1=225$ $r_g$ and $\xi_2=2700$ $r_g$, respectively, for which the broad and shallow red peak is best fit (figure 1).  This is important since, after subtraction of the disk and subsequent fitting of the emission lines (section 2.3), the [OIII]5007/[OIII]4959 ratio is $\sim$3, in agreement with atomic theory.  The correct position of the blue peak is also produced when these parameters are used.  However, in none of the fits does the symmetric disk profile account for all of the flux in the blue peak which is unusually strong for a double-peaked emitter.  A similar fit was performed on the H$\gamma$ region.  Though the H$\gamma$ emission is weaker and a fit is more difficult, it is clear that a symmetric disk profile is not sufficient there for the same reason that the blue H$\beta$ peak is not well modeled. 
   
\subsection{Emission Lines}
After subtraction of the disk profile, Gaussian profiles were fit to the broad and narrow central components of the H$\beta$/[OIII] and H$\gamma$ regions using the IRAF task 'fitprofs.'  In the H$\beta$/[OIII] region all parameters of the fit (peak position, peak flux and FWHM) are left as free parameters, but the peak position and FWHM in the H$\gamma$ region are fixed because the features are much weaker.  The FWHM of the narrow lines ([OIII]5007, [OIII]4959, H$\beta$, H$\gamma$, [NeIII]3869 and [OII]3727) range from 300 - 700 km s$^{-1}$ and probably originated in the same narrow line region (NLR).  The FWHM of all the broad central Balmer components (H$\beta$ and H$\gamma$) are also similar, indicating that they originated in a region nearer the central black hole, typical of a broad line region (BLR).  Gaussian profiles were also fit to the residuals of the blue H$\beta$ and H$\gamma$ peaks.  Table 1 lists the redshifts, Gaussian FWHMs and the line flux ratios relative to H$\beta$ for all of the fits.    
    
\section{Interpretation}
\subsection{Comparisons with Candidate Binary SMBHs}
While the asymmetry of the broad H$\beta$ component of SDSSJ0932+0318 that is blended with the [OIII] lines is likely the red shoulder of a broad double-peaked emitter, the strength of the blue H$\beta$ peak relative to the red peak is unusual for double-peaked emitters \citep{EH94}.  Additionally, the blue peak is offset from the central narrow line by $\sim$4200 km s$^{-1}$ which is on the high end for double-peaked emitters \citep[larger than 91 percent of the sample of 116 double-peaked emitters from][]{Strateva03}.  These properties warrant comparison with other objects identified from SDSS spectra that exhibit unusual double-peaked line profiles and which have been proposed as candidate binary SMBHs (figure 2).  

\emph{SDSSJ1536+0441:} The extended red wing and strong blue peak of SDSSJ0932+0318 is similar to the features seen in SDSSJ1536+0441.  The blue peak in SDSSJ1536+0441 is blueshifted by $\sim$3500 km s$^{-1}$, and the unusually high ratio of the blue peak to the broad, red peak makes it an extreme double-peaked emitter.  As such, SDSSJ1536+0441 has recently been proposed as both a double-peaked emitter and a binary SMBH \citep{TG09}.  This last hypothesis is particularly intriguing since it suggests that strong peaks in some double-peaked emitters may actually be caused by a second, smaller SMBH.  In this scenario, the less massive secondary SMBH is orbiting the primary SMBH and moving toward us, and the broad blue Balmer components in the spectrum may come from an accretion flow onto the secondary, or alternatively from the portions of the primary and circum-binary accretion disks near to and heated by the secondary \citep{TG09}.  

\emph{SDSSJ1050+3456:} A similarly extreme difference between the blue and red peaks is seen in SDSSJ1050+3456, and could likewise be interpreted as evidence for a double-peaked emitter in a binary SMBH.  Interestingly, however, the extended red wing is only apparent in the H$\alpha$ line of SDSSJ1050+3456, while the broad H$\beta$ peak appears to be symmetric.      

\emph{SDSSJ0927+2943:} The SDSS spectrum of SDSSJ0927+2943 is unique in that both the blue and red emission line systems contain narrow line components, including in the high ionization forbidden lines; and the broad components, which are only seen in the Balmer and MgII lines of the blue system, all appear symmetric with no visible disk profile.  If interpreted as a binary SMBH, the red system is associated with the NLR that surrounds both SMBHs and is at the redshift of the primary, more massive SMBH, and the blue system is associated with the secondary, smaller SMBH that provides illumination for both emission line systems.  

\begin{figure} $
\hspace*{-0.6in}
\begin{array}{cc}
\includegraphics[scale=.35,angle=270]{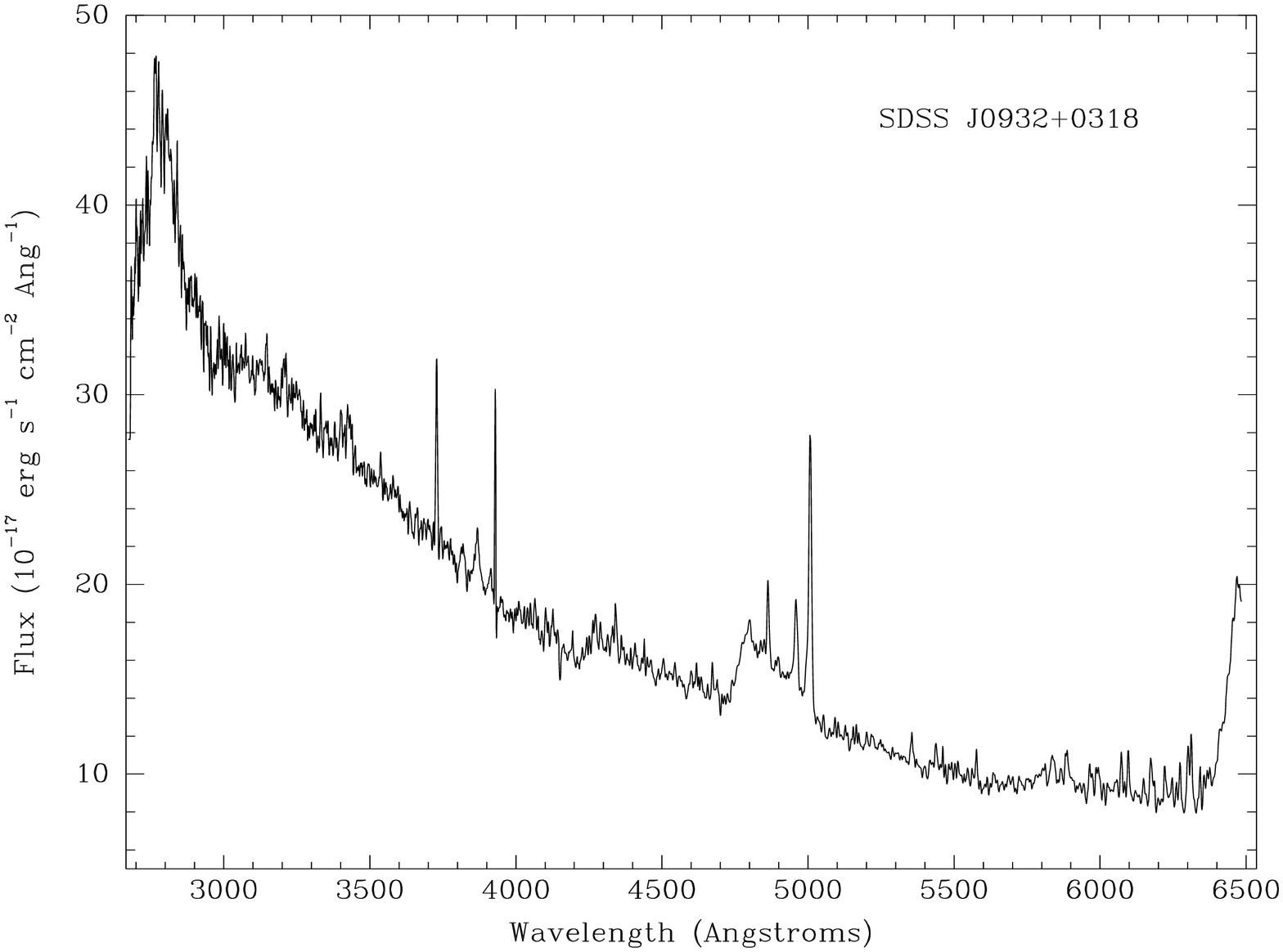} &
\includegraphics[scale=.35,angle=270]{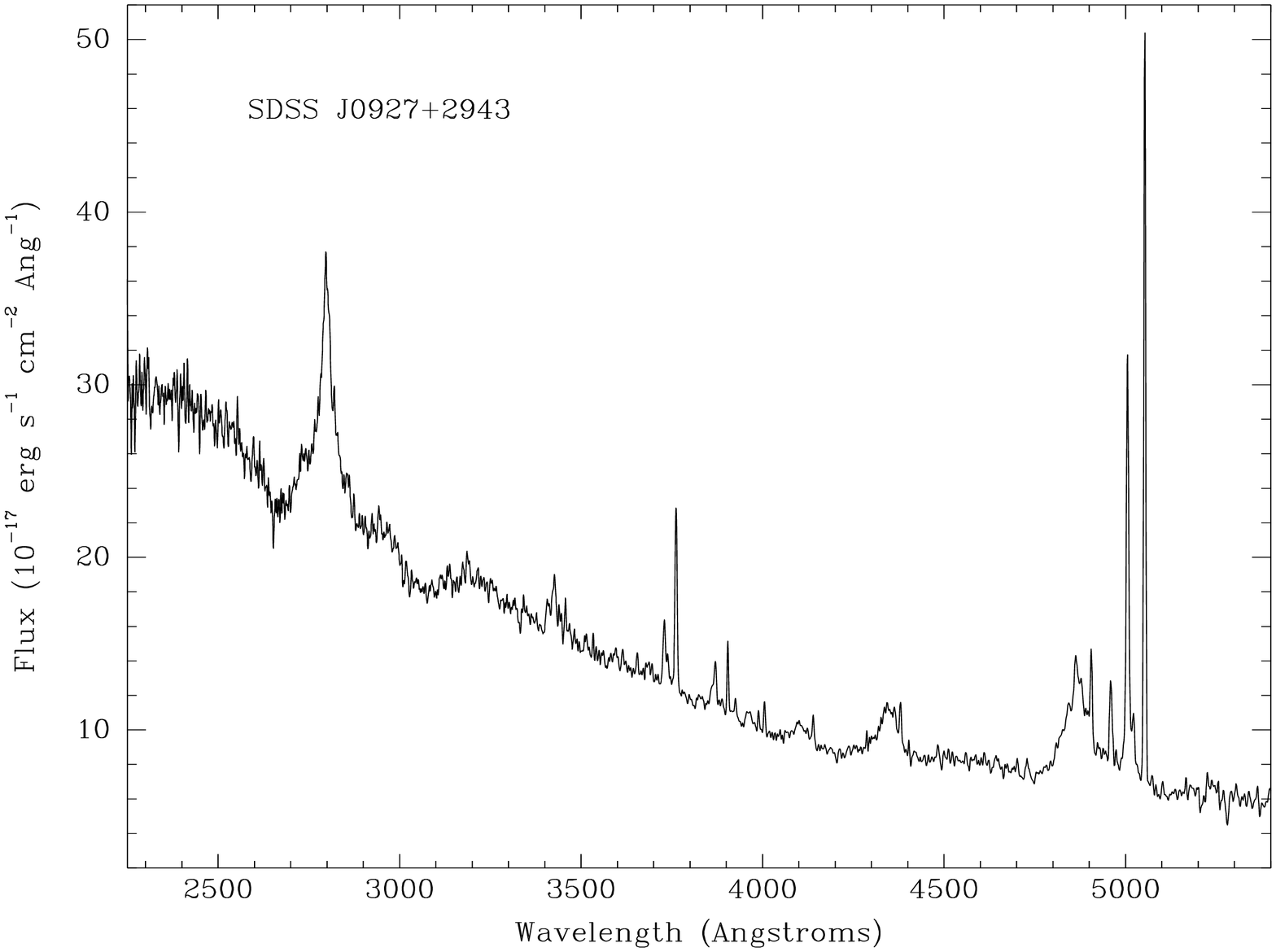} \\
\includegraphics[scale=.35,angle=270]{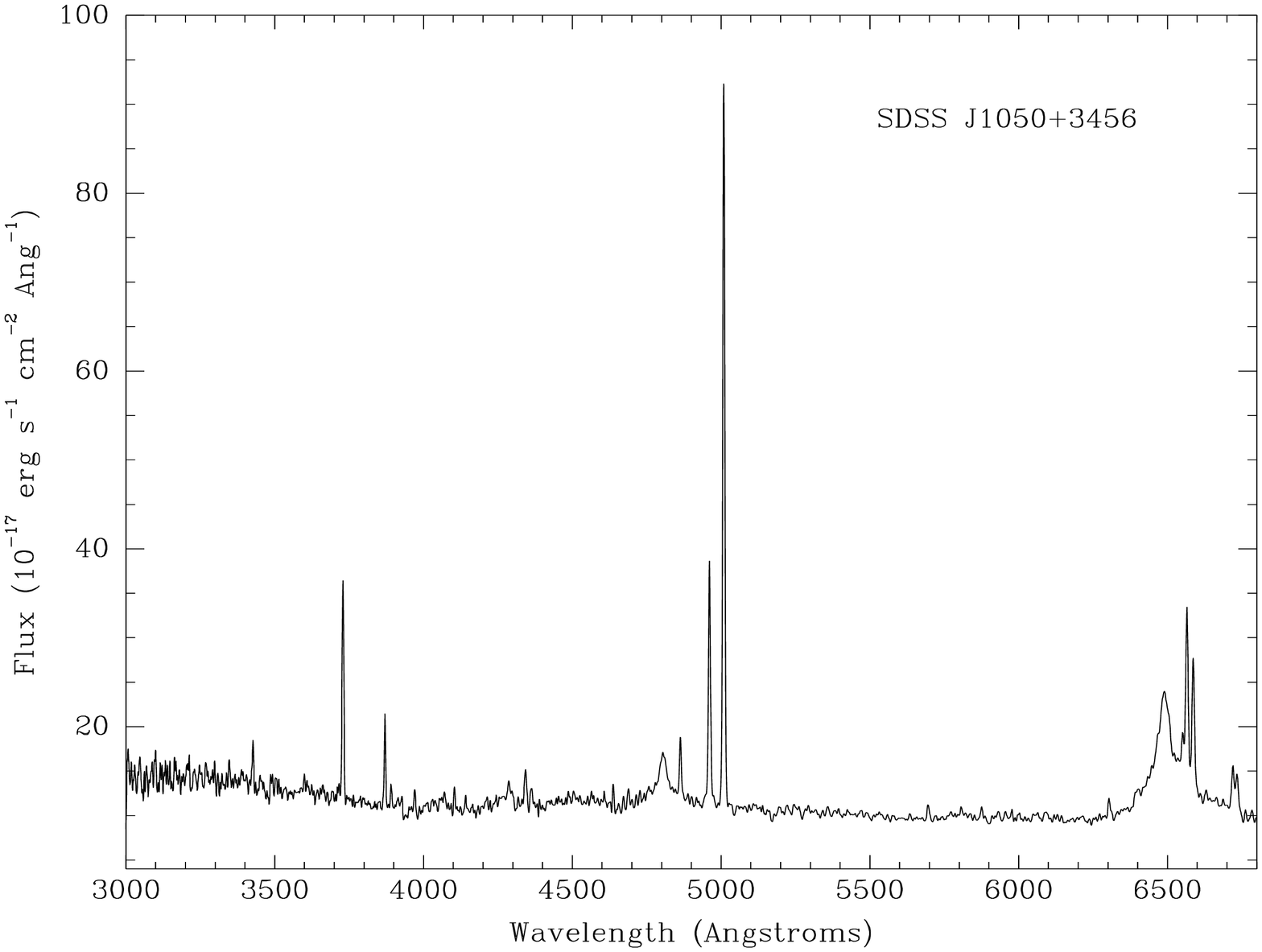} &
\includegraphics[scale=.35,angle=270]{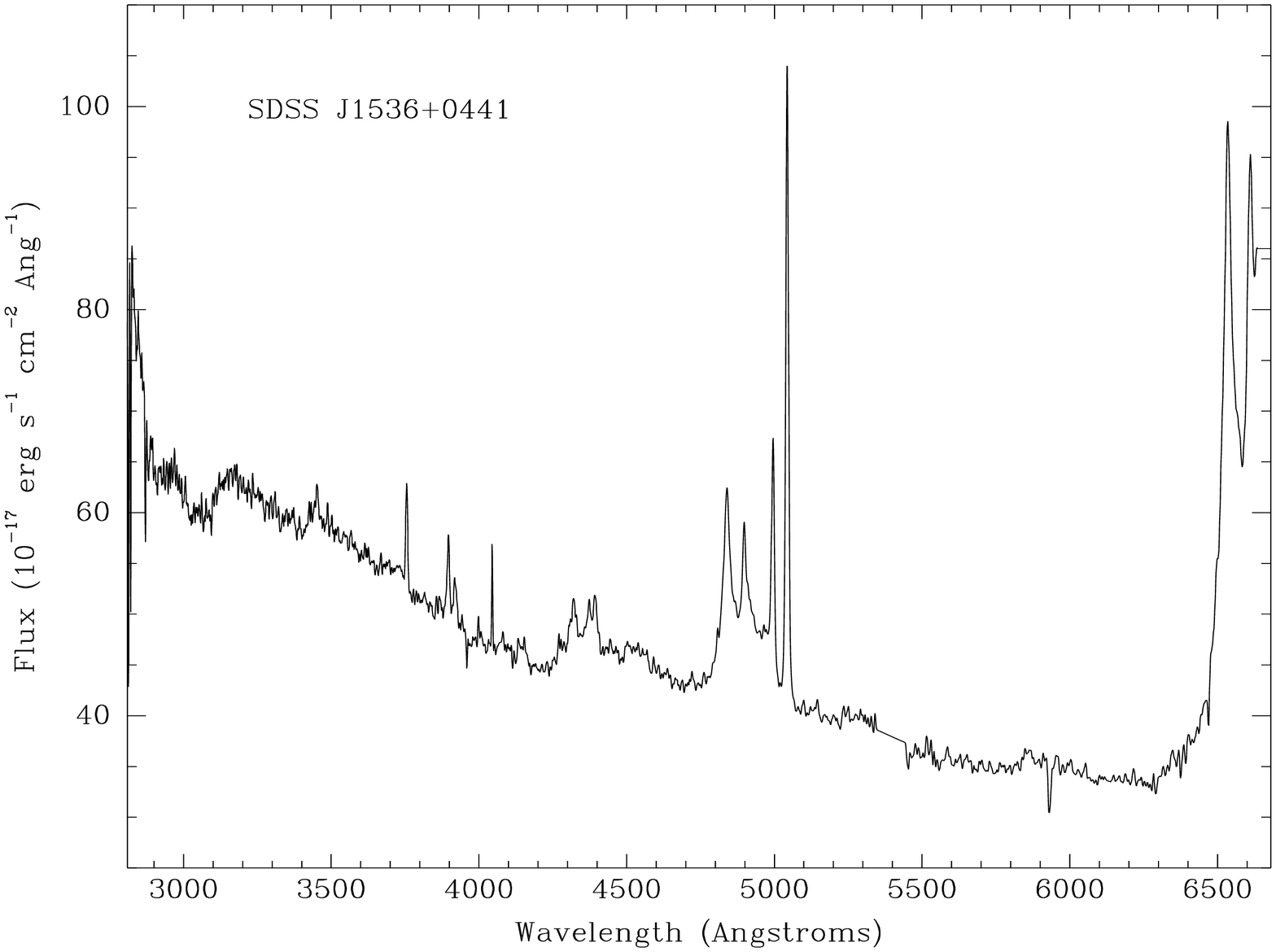}
\end{array} $
\caption{Top left: SDSS spectrum of SDSSJ0932+0318 with double-peaked H$\beta$ and H$\gamma$ emission lines.  Top right: SDSS spectrum of SDSSJ0927+2943 with double-peaked H$\beta$, H$\gamma$ and narrow emission lines.  Bottom right: SDSS spectrum of SDSSJ1050+3456 with double-peaked H$\alpha$, H$\beta$ and H$\gamma$ emission lines.  Bottom right: SDSS spectrum of SDSSJ1536+0441 with double-peaked H$\alpha$, H$\beta$ and H$\gamma$ emission lines \citep[the right side of H$\alpha$, which is cutoff in the SDSS spectrum, reveals a broad, shallow red wing ][]{Chornock10}.  All the spectra have been shifted to the rest-frame.  Notice that the double-peaks in the Balmer emission lines have varying degrees of strength and asymmetry.  For clarity, the spectra are smoothed to 600 km s$^{-1}$.}
\end{figure}   

\subsection{The Binary SMBH Model}
Testing these models on SDSSJ0932+0318 and other strongly double-peaked emission line profiles, is particularly important to do since spectra with double-peaked emission lines is a potential method of identifying binary SMBH candidates.   That the profile of a circular disk is inadequate suggests the possibility of an external influence, perhaps from a secondary SMBH.  Arguments in \citet{TG09} (and references therein) suggest that the presence of a secondary SMBH is effective at producing the strong peaks seen in some double-peaked emitters (and making a disk profile more visible by enhancing the illumination of the disk surface), especially those that require asymmetric disk profiles, and that many such double-peaked emitters may in fact be binary SMBHs.  

The models of potential interactions between binary SMBHs, accretion disks and circumbinary disks for the previously discovered candidates suggests that signatures of binary SMBHs are likely to be varied.  The visibility of a disk profile in the low-ionization lines is thought to be dependent on a number of factors: illumination (internal and external) of the geometrically thin portion of the disk, the disk radial size, and the orientation of the disk plane to the line of sight.  As for a second set of high ionization lines, arguments in \citet{Bogdanovic09b} and \citet{Dotti09} suggest that the secondary SMBH may be able to produce its own NLR by ionizing the low density region within the circumbinary disk.  The subject of our analysis, SDSSJ0932+0318, is clearly within the class of objects which show an apparent disk profile and for which there is no second set of high ionization lines.  In the binary SMBH scenario, this most likely implies that the secondary SMBH is not powerful enough to produce emission lines with high ionization potentials and low critical densities such as [OIII]5007, [OIII]4959, [NeIII]3869, [OII]3727 and [NeV]3426. The observed red NLR is the rarified medium that is large enough to surround two SMBHs that are separated by one parsec or less, and it is clearly at the redshift of the primary SMBH since it also coincides with the central wavelength of the disk model.  Following the model of \citet{TG09}, if the secondary SMBH is massive enough, then it may be able to maintain and illuminate its own BLR.  Alternatively, the secondary SMBH may tidally distort, or heat the outer edge of, the primary SMBH's accretion disk \citep{Lodato09}.  Both scenarios can explain the observed blue Balmer peaks (figure 3).     

\begin{figure}
\centerline{\epsfig{figure=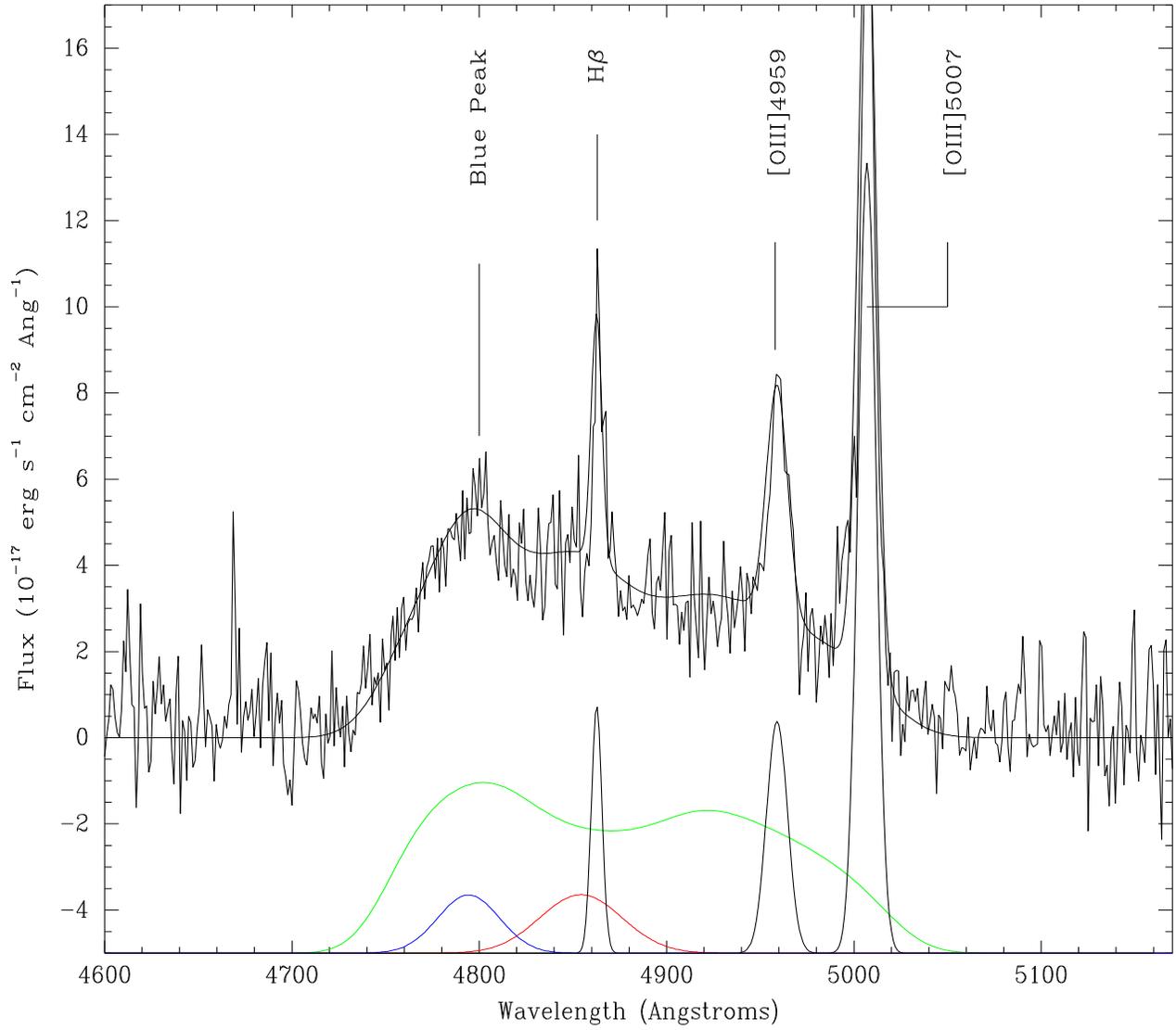,height=18cm,width=16cm,angle=270}}
\caption{H$\beta$/[OIII] complex at the rest wavelength with the disk profile plus best fit Gaussian components overlaid.  The Gaussian components were fit after subtraction of the disk model.  This analysis is similar to that of \citet{TG09}.}
\label{Hb_wdisk_emission}
\end{figure}

\begin{table}
\centering
\begin{tabular}{| l | c | c | c | }
\hline 
Emission Line & Redshift & FWHM (km s$^{-1}$) & Flux relative to $H\beta$ \\ \hline
\multicolumn{4}{| c |}{Central Narrow Components:} \\ \hline
[OIII]5007 & 0.4199 & 560 & 5.5 \\ \hline
[OIII]4959 & 0.4198 & 660 & 1.9 \\ \hline
H$\beta$ & 0.4201& 310 & 1.0 \\ \hline
H$\gamma$ & 0.4197 & ~320$^\ast$ & 0.5 \\ \hline
[NeIII]3869 & 0.4190 & 1800 & 2.4 \\ \hline
[OII]3727 & 0.4197 & 430 & 2.5 \\ \hline
\multicolumn{4}{| c |}{Central Broad Components:} \\ \hline
H$\beta$ & 0.4177 & 2710 & 1.0 \\ \hline
H$\gamma$ & 0.4179 & ~2710$^\ast$ & 0.3 \\ \hline
\multicolumn{4}{| c |}{Blue Components:} \\ \hline
H$\beta$ & 0.4001 & 2050 & 1.0 \\ \hline
H$\gamma$ & 0.3999 & ~2050$^\ast$ & 0.4 \\ \hline
\end{tabular} 
\caption{Redshifts, FWHMs and line ratios for the Gaussian fits to the spectrum.  The redshifts were calculated using the SDSS vacuum wavelengths.  The line widths were corrected for an instrumental resolution of 150 km s$^{-1}$.  $^\ast$Fixed values.}
\end{table}

\emph{Binary Parameters:}

Under the binary SMBH hypothesis, it is instructive to estimate the physical parameters of the system to determine if such a system may be identified from multiple observations and in what stage of binary SMBH evolution it would be placed.  The individual masses may be estimated if we make the assumption that the central and blue broad H$\beta$ components are from BLRs around the primary ($M_1$) and secondary ($M_2$) SMBHs, respectively.  In this case, FWHM$_{H\beta}$ and L$_{5100}$ may then be used with the relation of \citet{VP06}.  To determine the optical luminosities, the total bolometric luminosity calculated in section 2.1 (including the correction for 33 percent stellar contribution) was converted to a 5100\AA~ luminosity using $L_{5100}\approx L_{BOL}/9$ \citep{Peterson04}.  Then, the relative fraction contributed by each SMBH was calculated using the ratio of the H$\beta$ luminosities from the line fitting and converting to a ratio of the 5100\AA~ luminosities using the relation from \citet{GH05}.  The resulting luminosities are $L_{5100}=2.92\times10^{44}$ erg s$^{-1}$ and $L_{5100}=2.39\times10^{43}$ erg s$^{-1}$ for the primary and secondary SMBHs, respectively, providing mass estimates of $M_1\approx 3\times10^{7}$ M$_{\odot}$ and $M_2\approx 5\times10^{6}$ M$_{\odot}$ and a mass ratio of $M_2/M_1\approx 0.16$.  Additionally, using the combined mass, and assuming they are in a circular orbit with a relative velocity difference of ~3700 km s$^{-1}$ (the separation between the central and blue H$\beta$ broad components) the calculated separation is $\sim$0.01 pc or $\sim$14 l-d.  The velocity difference is very similar to that calculated by \citet{BL09}, but the separation is nearly an order of magnitude smaller because of the smaller total mass.  Though these are only rough estimates, they show that, if the system is a binary, it is in the final stage of the binary's lifetime.  

If the separation is small enough, there will likely be complex interactions between the secondary SMBH and the accretion disk of the primary SMBH.  We may estimate the physical disk size using the primary SMBH mass, which yields inner and outer radii of $\sim$0.0003 and $\sim$0.004 pc, respectively.  This result is interesting since it implies, based on the above calculations, that the secondary SMBH would be near the edge of the primary disk, potentially close enough to heat and/or distort it.  For the primary SMBH, we calculated the radius of the BLR \citep[from the relation of][]{Kaspi05} to be $\sim$0.008 pc or $\sim$9.6 l-d.  This result further implies that the binary system would be nearly inside of the primary's BLR.   However, the scaling relations using AGN luminosity may not directly apply to double-peaked emitters \citep{LE06} so these masses and orbital parameters are not necessarily realistic; they merely indicate that the sizes of accretion disks, BLRs and binary SMBH separations are of similar scales, allowing for complex interactions that are capable of producing the observed spectrum of many unusual double-peaked emitters.  For instance, the central broad components are slightly blueshifted ($\sim$500 km s$^{-1}$) relative to the central narrow components, indicating that there may be some sort of interaction between the secondary SMBH and the BLR of the primary SMBH.  In this case, the broad components will not be entirely due to virialized gas around the SMBHs.       
 
\subsection{Chance Projection}
Since the analysis of this quasar spectrum considers the possibility of a binary SMBH, we also consider the possibility of a chance projection of two quasars.  However, there is a small statistical likelihood of a positional coincidence.  Furthermore,  there are no blue components to the strong narrow forbidden lines ([OIII] and [OII]) that are typically seen in quasar spectra.  So, if the blue components in the spectrum are the result of a second SMBH, the second SMBH does not have a typical NLR and is more likely to be within the same NLR as the primary SMBH.

\subsection{Alternative Mechanisms for Asymmetry in the Disk}
In this section we investigate the possibility that only one SMBH is present and that the double peaks are produced entirely by a single accretion disk.  Based on the disk model fit in figure 1, this requires some form of asymmetry in the disk.  The structure of accretion disks around SMBHs has been widely studied, with various models proposed.  The model for the disk fitting we used \citep[from][]{CH89} has an inner torus photoionizing the geometrically thin disk that produces the disk line profile.  However, the illuminated portion of the disk may have a complex structure.  Studies of the variability of double-peaked emitters have shown that there may be large-scale patterns in the emissivity \citep{LE06}, suggesting that the surface emissivity is not uniform.  Proposed ideas for intrinsic asymmetries in accretion disks include hot spots \citep{Zheng91} and regions of high density such as spiral arms \citep{CW94} that may result in a region of higher luminosity that is in the blue-shifted portion of the disk.  Such an asymmetry would produce the unusually strong peaks seen in some double-peaked Balmer emission lines.  

\section{Conclusions}
In the case of SDSSJ0932+0318, the spectral features are unique and cannot be clearly classified as a typical double-peaked emitter.  While maintaining that a disk profile is indeed present in the Balmer emission lines, we explored the scenarios of a chance projection, a binary SMBH, and a single accretion disk with a complex and asymmetric  structure.  The chance projection is unlikely, especially since there is only one set of narrow [OIII] lines.  Since binary SMBHs are expected due to galaxy mergers, the model in which the strong blue peak in the disk is caused by a second SMBH is given serious consideration, especially since the velocity shift between the red and blue peaks in SDSSJ0932+0318 is similar to the shifts in the candidate binary SMBHs discovered previously and is on the high end for double peaked emitters.  However, since much about the physics of accretion disks around SMBHs is not fully understood, we cannot discount the possibility of intrinsic asymmetries in the disk.  

Perhaps one of the best and easiest tests of the binary hypotheses is multiple observations, which may reveal orbital motion if it exists.  From the basic binary parameters estimated in section 3.2, the hypothetical orbital period would be $\sim$10 years.  Such orbital motion may be apparent in the spectrum as a relative shifting of the two peaks over that time period, which would make it potentially observable.  It has been pointed out that follow-up observations of the three candidates previously mentioned (section 3.1) did not reveal a measurable velocity change \citep{Bogdanovic09a}.  However, it was also noted by \citet{Bogdanovic09a} that this does not negate the binary SMBH hypothesis but may merely imply that the projected change in velocity will only be detectable over a period of several years or more.  Even in the case where only a single SMBH is present, the strength of the blue H$\beta$ and H$\gamma$ emission peaks is notable and comparable to other ÒunusualÓ double-peaked emitters.  Understanding the nature of the complex line emission in this object will aid in further discrimination between a single SMBH with a complex accretion disk and the actual case of a binary SMBH.  Furthermore, the similarities of SDSSJ0932+0318 to other candidate binary SMBHs suggests that there may be a significant number of other objects with similar features that can be studied as binary SMBH candidates or unusual double-peaked emitters.      

\section*{Role of the Funding Source}
Support for this work was provided by the Arkansas NASA EPSCoR program (grant number NNX08AW03A).  

\section*{Acknowledgements}
We thank two anonymous reviewers for very helpful comments and suggestions that greatly improved the quality of the paper.  We would also like to acknowledge help from members of the Arkansas Galaxy Evolution Survey (AGES) and the Arkansas Center for Space and Planetary Sciences, Joel Berrier, Douglas Shields, Benjamin Davis, Jason Cuce, Adam Hughes and Jennifer Hanley, and also Daniel Stern, for helpful discussions.  This research has made use of NASA's Astrophysics Data System and the Sloan Digital Sky Survey.


\section*{References}

\begin{thebibliography}{30}
\expandafter\ifx\csname natexlab\endcsname\relax\def\natexlab#1{#1}\fi
\expandafter\ifx\csname url\endcsname\relax
  \def\url#1{\texttt{#1}}\fi
\expandafter\ifx\csname urlprefix\endcsname\relax\def\urlprefix{URL }\fi

\bibitem[{{Bogdanovi{\'c}} et~al.(2009{\natexlab{a}}){Bogdanovi{\'c}},
  {Eracleous}, and {Sigurdsson}}]{Bogdanovic09a}
{Bogdanovi{\'c}}, T., {Eracleous}, M., {Sigurdsson}, S., Jul.
  2009{\natexlab{a}}. {Emission lines as a tool in search for supermassive
  black hole binaries and recoiling black holes}. NewA 53, 113--120.

\bibitem[{{Bogdanovi{\'c}} et~al.(2009{\natexlab{b}}){Bogdanovi{\'c}},
  {Eracleous}, and {Sigurdsson}}]{Bogdanovic09b}
{Bogdanovi{\'c}}, T., {Eracleous}, M., {Sigurdsson}, S., May
  2009{\natexlab{b}}. {SDSS J092712.65+294344.0: Recoiling Black Hole or a
  Subparsec Binary Candidate?} ApJ 697, 288--292.

\bibitem[{{Bogdanovi{\'c}} et~al.(2008){Bogdanovi{\'c}}, {Smith}, {Sigurdsson},
  and {Eracleous}}]{Bogdanovic08}
{Bogdanovi{\'c}}, T., {Smith}, B.~D., {Sigurdsson}, S., {Eracleous}, M., Feb.
  2008. {Modeling of Emission Signatures of Massive Black Hole Binaries. I.
  Methods}. ApJ 174, 455--480.

\bibitem[{{Boroson} and {Lauer}(2009)}]{BL09}
{Boroson}, T.~A., {Lauer}, T.~R., Mar. 2009. {A candidate sub-parsec
  supermassive binary black hole system}. Nature 458, 53--55.

\bibitem[{{Chakrabarti} and {Wiita}(1994)}]{CW94}
{Chakrabarti}, S.~K., {Wiita}, P.~J., Oct. 1994. {Variable emission lines as
  evidence of spiral shocks in accretion disks around active galactic nuclei}.
  ApJ 434, 518--522.

\bibitem[{{Chen} and {Halpern}(1989)}]{CH89}
{Chen}, K., {Halpern}, J.~P., Sep. 1989. {Structure of line-emitting accretion
  disks in active galactic nuclei - ARP 102B}. ApJ 344, 115--124.

\bibitem[{{Chornock} et~al.(2010){Chornock}, {Bloom}, {Cenko}, {Filippenko},
  {Silverman}, {Hicks}, {Lawrence}, {Mendez}, {Rafelski}, and
  {Wolfe}}]{Chornock10}
{Chornock}, R., {Bloom}, J.~S., {Cenko}, S.~B., {Filippenko}, A.~V.,
  {Silverman}, J.~M., {Hicks}, M.~D., {Lawrence}, K.~J., {Mendez}, A.~J.,
  {Rafelski}, M., {Wolfe}, A.~M., Jan. 2010. {The Quasar SDSS J1536+0441: An
  Unusual Double-peaked Emitter}. ApJ 709, L39--L43.

\bibitem[{{Dotti} et~al.(2009){Dotti}, {Montuori}, {Decarli}, {Volonteri},
  {Colpi}, and {Haardt}}]{Dotti09}
{Dotti}, M., {Montuori}, C., {Decarli}, R., {Volonteri}, M., {Colpi}, M.,
  {Haardt}, F., Sep. 2009. {SDSSJ092712.65+294344.0: a candidate massive black
  hole binary}. MNRA 398, L73--L77.

\bibitem[{{Eracleous} and {Halpern}(1994)}]{EH94}
{Eracleous}, M., {Halpern}, J.~P., Jan. 1994. {Doubled-peaked emission lines in
  active galactic nuclei}. ApJS 90, 1--30.

\bibitem[{{Eracleous} and {Halpern}(2003)}]{EH03}
{Eracleous}, M., {Halpern}, J.~P., Dec. 2003. {Completion of a Survey and
  Detailed Study of Double-peaked Emission Lines in Radio-loud Active Galactic
  Nuclei}. ApJ 599, 886--908.

\bibitem[{{Gaskell}(1996)}]{Gaskell96}
{Gaskell}, C.~M., Jun. 1996. {Evidence for Binary Orbital Motion of a Quasar
  Broad-Line Region}. ApJ 464, L107+.

\bibitem[{{Greene} and {Ho}(2005)}]{GH05}
{Greene}, J.~E., {Ho}, L.~C., Sep. 2005. {Estimating Black Hole Masses in
  Active Galaxies Using the H{$\alpha$} Emission Line}. ApJ 630, 122--129.

\bibitem[{{Hopkins} et~al.(2005){Hopkins}, {Hernquist}, {Cox}, {Di Matteo},
  {Martini}, {Robertson}, and {Springel}}]{Hopkins05}
{Hopkins}, P.~F., {Hernquist}, L., {Cox}, T.~J., {Di Matteo}, T., {Martini},
  P., {Robertson}, B., {Springel}, V., Sep. 2005. {Black Holes in Galaxy
  Mergers: Evolution of Quasars}. ApJ 630, 705--715.

\bibitem[{{Jester} et~al.(2005){Jester}, {Schneider}, {Richards}, {Green},
  {Schmidt}, {Hall}, {Strauss}, {Vanden Berk}, {Stoughton}, {Gunn},
  {Brinkmann}, {Kent}, {Smith}, {Tucker}, and {Yanny}}]{Jester05}
{Jester}, S., {Schneider}, D.~P., {Richards}, G.~T., {Green}, R.~F., {Schmidt},
  M., {Hall}, P.~B., {Strauss}, M.~A., {Vanden Berk}, D.~E., {Stoughton}, C.,
  {Gunn}, J.~E., {Brinkmann}, J., {Kent}, S.~M., {Smith}, J.~A., {Tucker},
  D.~L., {Yanny}, B., Sep. 2005. {The Sloan Digital Sky Survey View of the
  Palomar-Green Bright Quasar Survey}. AJ 130, 873--895.

\bibitem[{{Kaspi} et~al.(2005){Kaspi}, {Maoz}, {Netzer}, {Peterson},
  {Vestergaard}, and {Jannuzi}}]{Kaspi05}
{Kaspi}, S., {Maoz}, D., {Netzer}, H., {Peterson}, B.~M., {Vestergaard}, M.,
  {Jannuzi}, B.~T., Aug. 2005. {The Relationship between Luminosity and
  Broad-Line Region Size in Active Galactic Nuclei}. ApJ 629, 61--71.

\bibitem[{{Komossa}(2006)}]{Komossa06}
{Komossa}, S., 2006. {Observational evidence for binary black holes and active
  double nuclei}. Memorie della Societa Astronomica Italiana 77, 733--+.

\bibitem[{{Komossa} et~al.(2008){Komossa}, {Zhou}, and {Lu}}]{Komossa08}
{Komossa}, S., {Zhou}, H., {Lu}, H., May 2008. {A Recoiling Supermassive Black
  Hole in the Quasar SDSS J092712.65+294344.0?} ApJ 678, L81--L84.

\bibitem[{{Lewis} and {Eracleous}(2006)}]{LE06}
{Lewis}, K.~T., {Eracleous}, M., May 2006. {Black Hole Masses of Active
  Galaxies with Double-peaked Balmer Emission Lines}. ApJ 642, 711--719.

\bibitem[{{Lodato} et~al.(2009){Lodato}, {Nayakshin}, {King}, and
  {Pringle}}]{Lodato09}
{Lodato}, G., {Nayakshin}, S., {King}, A.~R., {Pringle}, J.~E., Sep. 2009.
  {Black hole mergers: can gas discs solve the `final parsec' problem?} MNRAS
  398, 1392--1402.

\bibitem[{{Marconi} et~al.(2004){Marconi}, {Risaliti}, {Gilli}, {Hunt},
  {Maiolino}, and {Salvati}}]{Marconi04}
{Marconi}, A., {Risaliti}, G., {Gilli}, R., {Hunt}, L.~K., {Maiolino}, R.,
  {Salvati}, M., Jun. 2004. {Local supermassive black holes, relics of active
  galactic nuclei and the X-ray background}. MNRAS 351, 169--185.

\bibitem[{{Merritt} et~al.(2004){Merritt}, {Milosavljevi{\'c}}, {Favata},
  {Hughes}, and {Holz}}]{MM04}
{Merritt}, D., {Milosavljevi{\'c}}, M., {Favata}, M., {Hughes}, S.~A., {Holz},
  D.~E., May 2004. {Consequences of Gravitational Radiation Recoil}. ApJ 607,
  L9--L12.

\bibitem[{{Peterson} et~al.(2004){Peterson}, {Ferrarese}, {Gilbert}, {Kaspi},
  {Malkan}, {Maoz}, {Merritt}, {Netzer}, {Onken}, {Pogge}, {Vestergaard}, and
  {Wandel}}]{Peterson04}
{Peterson}, B.~M., {Ferrarese}, L., {Gilbert}, K.~M., {Kaspi}, S., {Malkan},
  M.~A., {Maoz}, D., {Merritt}, D., {Netzer}, H., {Onken}, C.~A., {Pogge},
  R.~W., {Vestergaard}, M., {Wandel}, A., Oct. 2004. {Central Masses and
  Broad-Line Region Sizes of Active Galactic Nuclei. II. A Homogeneous Analysis
  of a Large Reverberation-Mapping Database}. ApJ 613, 682--699.

\bibitem[{{Schlegel} et~al.(1998){Schlegel}, {Finkbeiner}, and
  {Davis}}]{Schlegel98}
{Schlegel}, D.~J., {Finkbeiner}, D.~P., {Davis}, M., Jun. 1998. {Maps of Dust
  Infrared Emission for Use in Estimation of Reddening and Cosmic Microwave
  Background Radiation Foregrounds}. ApJ 500, 525--+.

\bibitem[{{Schneider} et~al.(2007){Schneider}, {Hall}, {Richards}, {Strauss},
  {Vanden Berk}, {Anderson}, {Brandt}, {Fan}, {Jester}, {Gray}, {Gunn},
  {SubbaRao}, {Thakar}, {Stoughton}, {Szalay}, {Yanny}, {York}, {Bahcall},
  {Barentine}, {Blanton}, {Brewington}, {Brinkmann}, {Brunner}, {Castander},
  {Csabai}, {Frieman}, {Fukugita}, {Harvanek}, {Hogg}, {Ivezi{\'c}}, {Kent},
  {Kleinman}, {Knapp}, {Kron}, {Krzesi{\'n}ski}, {Long}, {Lupton}, {Nitta},
  {Pier}, {Saxe}, {Shen}, {Snedden}, {Weinberg}, and {Wu}}]{Schneider07}
{Schneider}, D.~P., {Hall}, P.~B., {Richards}, G.~T., {Strauss}, M.~A., {Vanden
  Berk}, D.~E., {Anderson}, S.~F., {Brandt}, W.~N., {Fan}, X., {Jester}, S.,
  {Gray}, J., {Gunn}, J.~E., {SubbaRao}, M.~U., {Thakar}, A.~R., {Stoughton},
  C., {Szalay}, A.~S., {Yanny}, B., {York}, D.~G., {Bahcall}, N.~A.,
  {Barentine}, J., {Blanton}, M.~R., {Brewington}, H., {Brinkmann}, J.,
  {Brunner}, R.~J., {Castander}, F.~J., {Csabai}, I., {Frieman}, J.~A.,
  {Fukugita}, M., {Harvanek}, M., {Hogg}, D.~W., {Ivezi{\'c}}, {\v Z}., {Kent},
  S.~M., {Kleinman}, S.~J., {Knapp}, G.~R., {Kron}, R.~G., {Krzesi{\'n}ski},
  J., {Long}, D.~C., {Lupton}, R.~H., {Nitta}, A., {Pier}, J.~R., {Saxe},
  D.~H., {Shen}, Y., {Snedden}, S.~A., {Weinberg}, D.~H., {Wu}, J., Jul. 2007.
  {The Sloan Digital Sky Survey Quasar Catalog. IV. Fifth Data Release}. AJ
  134, 102--117.

\bibitem[{{Shields} et~al.(2009){Shields}, {Rosario}, {Smith}, {Bonning},
  {Salviander}, {Kalirai}, {Strickler}, {Ramirez-Ruiz}, {Dutton}, {Treu}, and
  {Marshall}}]{Shields09}
{Shields}, G.~A., {Rosario}, D.~J., {Smith}, K.~L., {Bonning}, E.~W.,
  {Salviander}, S., {Kalirai}, J.~S., {Strickler}, R., {Ramirez-Ruiz}, E.,
  {Dutton}, A.~A., {Treu}, T., {Marshall}, P.~J., Dec. 2009. {The Quasar SDSS
  J105041.35+345631.3: Black Hole Recoil or Extreme Double-Peaked Emitter?} ApJ
  707, 936--941.

\bibitem[{{Strateva} et~al.(2003){Strateva}, {Strauss}, {Hao}, {Schlegel},
  {Hall}, {Gunn}, {Li}, {Ivezi{\'c}}, {Richards}, {Zakamska}, {Voges},
  {Anderson}, {Lupton}, {Schneider}, {Brinkmann}, and {Nichol}}]{Strateva03}
{Strateva}, I.~V., {Strauss}, M.~A., {Hao}, L., {Schlegel}, D.~J., {Hall},
  P.~B., {Gunn}, J.~E., {Li}, L., {Ivezi{\'c}}, {\v Z}., {Richards}, G.~T.,
  {Zakamska}, N.~L., {Voges}, W., {Anderson}, S.~F., {Lupton}, R.~H.,
  {Schneider}, D.~P., {Brinkmann}, J., {Nichol}, R.~C., Oct. 2003.
  {Double-peaked Low-Ionization Emission Lines in Active Galactic Nuclei}. AJ
  126, 1720--1749.

\bibitem[{{Tang} and {Grindlay}(2009)}]{TG09}
{Tang}, S., {Grindlay}, J., Oct. 2009. {The Quasar SDSS J153636.22+044127.0: A
  Double-Peaked Emitter in a Candidate Binary Black Hole System}. ApJ 704,
  1189--1194.

\bibitem[{{Vestergaard} and {Peterson}(2006)}]{VP06}
{Vestergaard}, M., {Peterson}, B.~M., Apr. 2006. {Determining Central Black
  Hole Masses in Distant Active Galaxies and Quasars. II. Improved Optical and
  UV Scaling Relationships}. ApJ 641, 689--709.

\bibitem[{{Zheng} et~al.(1991){Zheng}, {Veilleux}, and {Grandi}}]{Zheng91}
{Zheng}, W., {Veilleux}, S., {Grandi}, S.~A., Nov. 1991. {3C 390.3 - Modeling
  variable profile humps}. ApJ 381, 418--425.

\bibitem[{{Zhou} et~al.(2004){Zhou}, {Wang}, {Zhang}, {Dong}, and
  {Li}}]{Zhou04}
{Zhou}, H., {Wang}, T., {Zhang}, X., {Dong}, X., {Li}, C., Mar. 2004. {Obscured
  Binary Quasar Cores in SDSS J104807.74+005543.5?} ApJ 604, L33--L36.

\end{thebibliography}

\end{document}